\documentclass[journal, letterpaper, twocolumn,a4paper]{IEEEtran}
\IEEEoverridecommandlockouts
\usepackage{cite}
\usepackage{amsmath,amssymb,amsfonts}
\usepackage{dsfont}
\usepackage[latin9]{inputenc}
\usepackage{graphicx}
\usepackage{algpseudocode}
\usepackage{textcomp}
\usepackage{bbm}
\usepackage{xcolor}
\usepackage{booktabs}
\usepackage{subcaption}
\usepackage{dblfloatfix}
\usepackage{soul} 
\usepackage{comment}
\newcommand{\comm}[1]{}
\usepackage[ruled,vlined,linesnumbered]{algorithm2e}
\usepackage{multirow}
\def\BibTeX{{\rm B\kern-.05em{\sc i\kern-.025em b}\kern-.08em
    T\kern-.1667em\lower.7ex\hbox{E}\kern-.125emX}}
\usepackage{amsmath}
\DeclareMathOperator*{\argmax}{arg\,max}
\newtheorem{remark}{Remark}
\newcommand{\red}{\textcolor{red}}

\newcommand{\blue}{\textcolor{blue}}
\newcommand{\adapt}{AdaptSky}

\newcommand{\los}{LoS}

\newcommand{\mmwave}{mmWave}
\newcommand{\uav}{\ensuremath {\tt{UAV}}}
\newcommand{\uavs}{\ensuremath {\tt{UAV}s}}
\newcommand{\mwave}{\ensuremath {\tt{mmWave}}}
\newcommand{\nue}{\ensuremath {\tt{N_{UE}}}}
\newcommand{\ue}{\ensuremath {\tt{UE}}}
\newcommand{\mimo}{\ensuremath {\tt{MIMO}}}
\newcommand{\subs}{\ensuremath {\tt{sub-6}}}
\newcommand{\nlosm}{\ensuremath {\tt{NLoS}}}
\newcommand{\losm}{\ensuremath {\tt{LoS}}}

\newcommand{\sub}{sub-6GHz}

 \usepackage{geometry}

\begin{document}

\title{
  \adapt: 
A DRL Based Resource Allocation Framework in NOMA-UAV Networks}

\author{\IEEEauthorblockN{ Ahmed Benfaid$^*$, Nadia Adem$^*$, and Bassem Khalfi$^{**}$}\\
\IEEEauthorblockA{$^*$University of Tripoli, Tripoli, Libya, E-mail: \{a.benfaid,n.adem\}@uot.edu.ly}\\
\IEEEauthorblockA{$^{**}$Qualcomm Technologies Inc., CA, USA, Email: bkhalfi@qti.qualcomm.com\\
} 
}

\maketitle   
\begin{abstract} 
Unmanned aerial vehicle (\uav) has recently attracted a lot of attention as a candidate to meet the 6G ubiquitous connectivity demand and boost the resiliency of terrestrial networks. Thanks to the high spectral efficiency and low latency, non-orthogonal multiple access (NOMA) is a potential access technique for future communication networks.
In this paper, we propose to use the \uav~as a moving base station (BS) to serve multiple users using NOMA and  jointly solve for  the 3D-\uav~placement and resource allocation  problem. Since the corresponding optimization problem is non-convex,  
we rely on the recent advances in artificial intelligence (AI) and propose \adapt, a deep reinforcement learning (DRL)-based framework, to efficiently solve it. 
To the best of our knowledge, \adapt~is the first framework that optimizes NOMA power allocation jointly with 3D-\uav~placement using both \sub~and millimeter wave mmWave spectrum.  Furthermore, for the first time in NOMA-UAV networks,  \adapt~integrates the dueling network (DN) architecture to the DRL technique to improve its learning capabilities. Our findings show that \adapt~does not only exhibit a fast-adapting 
learning and outperform the state-of-the-art baseline approach in data rate and fairness, but also it generalizes very well. The AdaptSky source code is accessible to use here: https://github.com/Fouzibenfaid/AdaptSky 
\end{abstract}

 \vspace{-1em}
\section{Introduction} 
\label{sec: Introd}
Future communication networks are envisioned to provide heterogeneous wireless communication services with significant performance boost over 5G networks~\cite{dang2020should}.
 Unmanned aerial vehicles (\uavs), artificial intelligence (AI), millimeter wave (mmWave), along with some emerging medium access techniques are very key technologies in meeting such a goal~\cite{letaief2019roadmap}.
For instance, thanks to their flexible 3D mobility, ease of deployment, and location precision, \uavs~can serve as aerial base stations (BSs), and hence, augment or replace  terrestrial BSs in some extreme scenarios~\cite{chowdhury20206g}. Hence, it has become an active topic for different working groups in standardization bodies such as 3GPP~\cite{lin2018sky}. 
%

 Rendering to its ability to simultaneously share spectrum resources among multiple users, non-orthogonal multiple access (NOMA)  promises for massive-devices connectivity making it a candidate access technique for beyond 5G systems. 
 Moreover, when used in mmWave spectrum, NOMA offers a significant data rate boost.
Nevertheless, efficiently managing resources for mmWave or even \sub~spectrum for NOMA-\uav~networks with a heterogeneous number of users is a complex problem~\cite{liu2019uav}. 
In this perspective, we propose a novel advanced deep reinforcement learning (DRL)-based framework that solves jointly for the \uav~placement and NOMA resource allocation in the context of \sub~as well as  \mmwave~spectrum.
%
\label{subsec: related}

There are few works available about NOMA-\uav~networks, some of which have   focused  on \uav~placement and NOMA resources allocation~\cite{Chen2019,Sharma2017,cui2018joint,Monemi2020}. 
%
The issues, however, with these works that, for simplicity they  $i)$ assume links between users and \uavs~are dominated by  line-of-sight (LoS),  as it is the case with~\cite{cui2018joint,Sharma2017,Chen2019}, 
$ii)$ restrict the number of users in the network to two, e.g.~\cite{Monemi2020,Sharma2017}, 
 $iii)$ solve for  \uav~placement and NOMA power allocation  disjointly, for example~\cite{Chen2019}, 
 and/or 
$iv)$ limit the \uav~analysis to 2D placement,  e.g.~\cite{Chen2019,Monemi2020,Sharma2017,cui2018joint}.
There have  been  some attempts in handling the 3D-placement problem of \uav~such as in~\cite{el2019learn,Alzenad2017,Yaliniz2016}. Yet, these works disjointly optimize  the altitude and the 2D-\uavs~placement. Imposing restrictions in analyzing 3D networks,  
can lead to inefficient use of resources. Hence,  some innovative techniques that can handle  their sophisticated analysis are  needed. 

A remarkable success has been reported for AI from incorporating DRL  into the field of gaming~\cite{deepmind2013}. DRL, which is mainly a reinforcement learning (RL) technique 
 combined with deep-neural-network (DNN), has the potential to handle high-dimensional inputs, learn patterns, and solve  complex problems efficiently~\cite{deepmind2013,mnih2015human}. It has recently witnessed a number of advances to improve DRL learning abilities, speed, and generalization.
 Even though some initial works consider DRL for  2D-\uavs~placement 
  in the presence of \los~links only,  e.g.~\cite{liu2018energy}, which employs deep deterministic policy gradient (DDPG) DRL, the  full potentials of DRL for 3D networks still need to be assessed.
 Integrating dueling network architectures with DRL 
   has shown its merits in dramatically improving and generalizing  learning in the   Atari domain~\cite{wang2016dueling}. 
   In the same regard, dueling DRL,  compared to other DRL advances like double DRL, as discussed in the \uav~sensing related application context~\cite{kersandt2018self}, shows an improvement in the learning speed. Nevertheless, there are  surprisingly very limited research investigations done about integrating  dueling DRL and~\uav.
 We aim, in this paper, to 
solve the non-convex optimization problem of the 3D-network resources management using our proposed dueling DRL based framework. 

%
\label{subsec: Contri}
In this paper, we aim to bring forth the most-recent advances in AI to efficiently solve the \uav~placement and resources management
while maximizing both users\textquotesingle~data rate and fairness. Our main contributions are summarized as follows 
\begin{enumerate}
   \item We propose a unified model-free framework,  \adapt, for 3D \uav~placement and power allocation in a NOMA-based network.  We integrate dueling network with DRL, for the first time in NOMA-UAV networks, and demonstrate the tremendous-gain resultant in learning model generalization, and hence,  network performance. 
  \item We show that \adapt~is robust for different channel gain models  both for the \sub~and mmWave spectrum.
 \item We show that \adapt~maximizes the spectral efficiency while maintaining high users\textquotesingle~fairness. Simulation results shows that \adapt~outperforms other optimization based mathematical  framework approach. 
  \end{enumerate}

\comm{\subsection{Roadmap}
 \label{subsec: Organiz}
 The rest of this paper is organized as follows. The system model and problem formulation are presented in Section \ref{sec: Sys Model} and  \ref{sec: Problem}, respectively. \adapt~framework is proposed and discussed in Section \ref{sec:adaptsky}. In Section \ref{sec: numr}, we provide the simulation results. 
Finally,  we present the conclusion in Section
\ref{sec: conc}. 
\vspace{-1.5em}
}
\section{System model}
\label{sec: Sys Model}

\begin{figure}[t]
    \centering
    \includegraphics[width=0.4\textwidth]{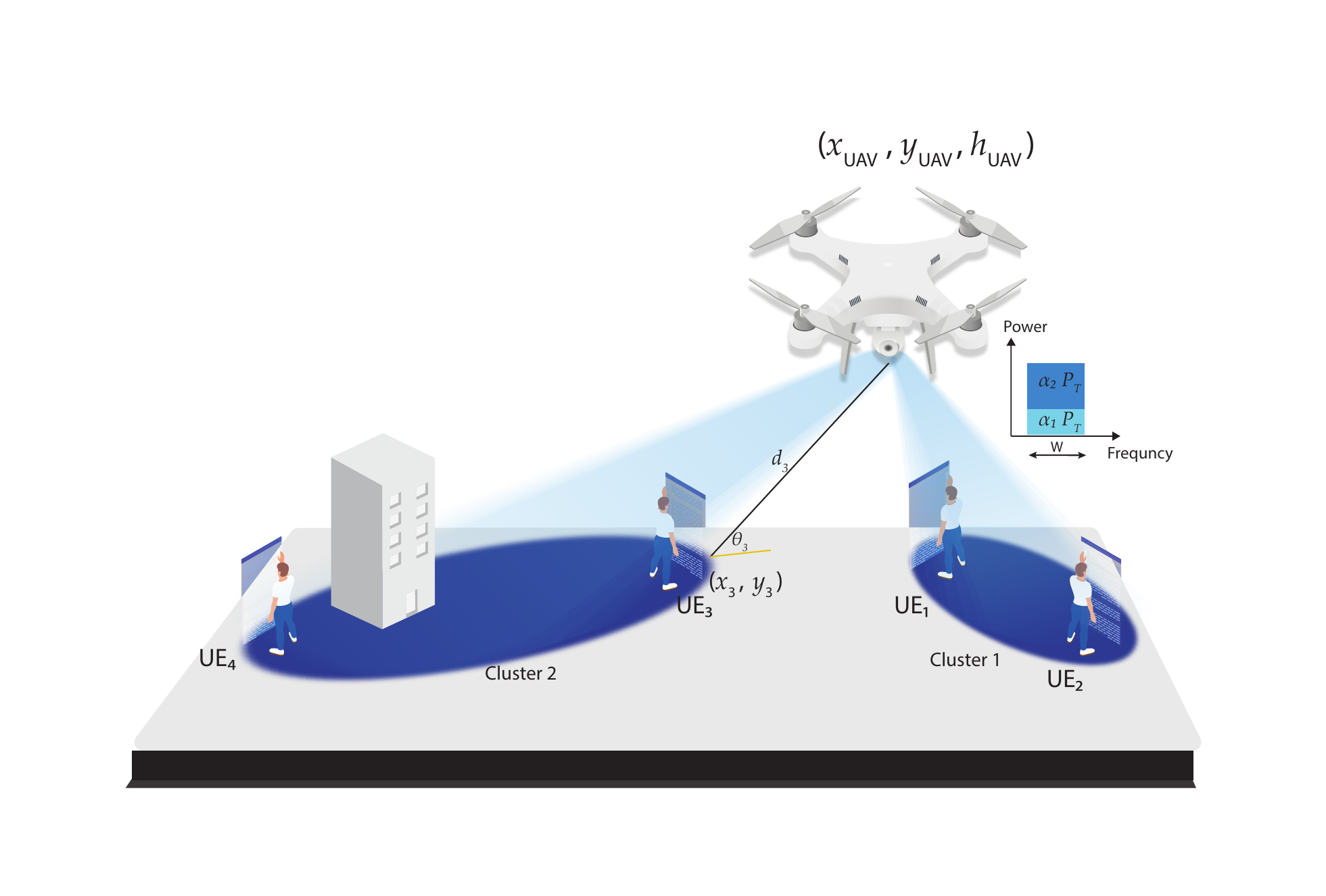}
    \caption{System model.}
    \label{fig:system_model}
\end{figure}


\subsection{Network Model} 
\label{subsec: Network} 
We consider a downlink cellular network with a \uav~serving $ \nue=2 {N}$ ground users distributed randomly over an area $\mathcal{A}$ of $L \times L$ units and grouped into a number  {of}  clusters.     
To get the most out of NOMA,  two users with  {distinct} channel qualities get associated to a single cluster based on the strategy proposed in~\cite{liu2019uav}.  
The \uav~serves each cluster over an orthogonal resource with  a total power $P_T$  distributed between the two corresponding users. The \uav~is also assumed to be  equipped with $\mathcal{N}_{\uav}$ antennas while each user is equipped with $\mathcal{N}_{\ue}$ antennas, unless specified otherwise. 
Throughout the paper, user $i$ is denoted by ${\ue}_i$ where $i \in \{1,2,..,\nue\}$.  Without loss of generality, we assume that  users ${\ue}_i$  and ${\ue}_{i+1}$  for $i \in \{1,3,..,\nue-1 \}$  are associated with the same cluster and 
 $\ue_i$ has a stronger channel gain than ${\ue}_{i+1}$.
 The received power 
at the ${\ue}_i$  
at a given time  step $\tau$ can be expressed as
\begin{eqnarray} \label{eqn: output}
 \hat{P}_{i,\tau}  &=& P_T{g}_{i,\tau}^{\mimo}(d_{i,\tau})  {\alpha}_{i,\tau} ,
\end{eqnarray}
where ${g}^{\mimo}_{i,\tau}(d_{i,\tau})$ is the channel gain 
between the \uav~and ${\ue}_i$  separated, at time $\tau$, by a 3D distance $d_{i,\tau}$. Considering only a large scale fading, small scale fading is deferred for future works,  and assuming a slight difference between antenna pairs, the channel gain can be approximated as ${g}^{\mimo}_{i,\tau}(d_{i,\tau}) = G {g}_{i,\tau}(d_{i,\tau})$, where  $G=\mathcal{N}_{\uav}\times \mathcal{N}_{\ue}$.  ${g}_{i,\tau}(d_{i,\tau})$ is the channel gain between one UAV-${\ue}_i$ antenna pair. ${\alpha}_{i,\tau}$ is the power allocation coefficient  that determines the amount of power, out of $P_T$, the \uav~assigns to ${\ue}_i$ at the time step $\tau$. 
%
\subsection{Signal-to-Interference-plus-Noise Ratio (SINR)}
\label{subsec: SINR}
Following the NOMA protocol,  the superposition coding (SC) is used  at the \uav~to transmit  messages for  users located in the same cluster. SC encodes different messages into a single signal while assigning them different power values.
 The successive interference cancellation (SIC) is used at the receiver side for signal detection. The received SINR at ${\ue}_i$ is expressed as 
\begin{equation} \label{eqn: SINR_far}
SINR_{i,\tau}  = \frac{P_T \times {g}_{i,\tau}(d_{i,\tau}) \times G \times\alpha_{i,\tau} }{P_T \times {g}_{i,\tau}(d_{i,\tau}) \times G \times \beta_{i,\tau}  + \sigma^2},
\end{equation}
%
where $\beta_{i,\tau}=\alpha_{i-1,\tau}$ if $i$ is even and zero otherwise.
  $\sigma^2$ is the noise power.
  The first term in the denominator of equation (\ref{eqn: SINR_far}) represents the interference from the user with the best channel condition to the other user. Based on the SIC technique, however, the interference at  the user with the best channel condition gets canceled, which is the intuition behind the definition of $\beta_{i,\tau}$.     
For a bandwidth $W$, the data rate of ${\ue}_i$ at $\tau$ is given by
\begin{equation} \label{eqn: sum_rate}
R_{i,\tau}  = W\log_2(1+SINR_{i,\tau} ).
\end{equation}
\vspace{-2em}
\subsection{Channel Model}
\label{subsec: Channel} 
We consider to explore the performance of our system in both mmWave and \sub~spectrum in the presence of the LoS and NLoS links. 
 To accommodate for the various links and spectrum technologies,   we modify the channel gain notation 
  to be ${g^{k,sp}_{i,\tau}}(d_{i,\tau})$, where $k \in \{\losm, \nlosm\}$ and $sp \in \{\mwave,\subs\}$. 
\subsubsection{MmWave Channel Model} Following the same model as in~\cite{al2014optimal}, at time step $\tau$, the \uav-$\ue_i$ link is assumed to be in LoS with a probability $Pr^{\losm,\mwave}_{i,\tau}(\theta_{i,\tau})$ given as  
\begin{equation} \label{eqn: mmWave-prob}
 Pr^{\losm,\mwave}_{i,\tau}(\theta_{i,\tau}) = \frac{1}{1 + C \exp[-Y  (\theta_{i,\tau}   \frac{180}{\pi} - C)]},
\end{equation}
where $C$ and $Y$ are environment parameters, $\theta_{i,\tau}$ is the elevation angle   between the \uav~and $\ue_i$.
Similarly, $\ue_i$ is assumed to be in NLoS using the complement property. 
The channel gain equation ${g}^{k,\mwave}_{i,\tau}(d_{i,\tau})$ between the \uav~located at a distance $d_{i,\tau}$ from $\ue_i$, 
according to~\cite{akdeniz2014millimeter} is expressed as
\begin{equation} \label{eqn: mmWave-pl}
 {g}^{k,\mwave}_{i,\tau}(d_{i,{\tau}}) = C_k d_{i,\tau}^{-a_k},
\end{equation}
where $a_k$ is the path loss exponent, and $C_k$ is the unit distance path loss.
\subsubsection{Sub-6GHz Channel Model} Similar to~\cite{al2014modeling}, $\ue_i$ is assumed to be in LoS with a probability given by 
\begin{equation} \label{eqn: sub-6 GHz-prob}
Pr^{\losm,\subs}_{i,\tau}(\theta_{i,\tau}) = C.(\theta_{i,\tau} - \theta_0)^Y,
\end{equation}
where $C$ and $Y$ are frequency and environment dependent parameters. 
$\theta_0$ is the minimum angle allowed by the model.  Similarly, $\ue_i$ is assumed to be in NLoS using the complement property $Pr^{\nlosm,\subs}_{i,\tau}(\theta_{i,\tau}) = 1 - Pr^{\losm,\subs}_{i,\tau}(\theta_{i,\tau})$.
 The channel gain model for the $\ue_i$ is  defined as
\begin{equation} \label{eqn: sub-6 GHz-pl}
{g}_{i,\tau}^{k,\subs}(d_{i,\tau}) =   \big(\frac{c}{4 \pi f_c d_{i,\tau}}\big)^2 10^{ -0.1\eta_{k}},
\end{equation}
where $f_c$ is the carrier frequency, and 
$(\frac{c}{4 \pi f_c d_{i,\tau}}\big)^2$
represents the free space path loss. $ \eta_{k} $, measured in dB, is the mean additional loss for  
transmission link~\cite{al2014modeling}.

\subsection{\uav~Mobility Model}
%
At time step $\tau$,  the \uav~is assumed to be placed at $(x_{\uav,\tau},y_{\uav,\tau},h_{\uav,\tau})$ and   able to move to %
 $(x_{\uav,\tau}+d_x \delta_x , y_{\uav,\tau}+d_y \delta_y , h_{\uav,\tau}+ d_h \delta_h)$, where  $d_{x}$, $d_{y}$, and $d_{h}$ $\in\{1,-1\}$. 
 $\delta_x$,  $\delta_y$, and $\delta_h$  are the magnitude of change  in the $x$ and $y$ axis, and height, respectively.  $ h_{\uav,\tau} \geq h_0$, by assumption. 
 We assume that at  {$\tau=0$}, the \uav~is located at $(0,0,h_{\uav,0})$, where $h_{\uav,{0}}$ is the initial height. 
 %
%
%
Without loss of generality, we assume the \uav~can collect the channel state information (CSI) at the beginning of each time step~\cite{el2016power}.\\
The system model is shown in  Fig.~\ref{fig:system_model}. For simplicity,   $\tau$ is dropped for system parameters in the figure. 

\section{3D-\uav~Placement and Power Allocation Formulation} 
  \label{sec: Problem} 
We propose to optimize the \uav~placement and power allocation that maximizes the total users\textquotesingle~data rate and fairness. 
First, we define the sum users\textquotesingle~data rate at time step $\tau$ as

\begin{equation}
    R_{\tau}^{\textrm{tot}} =  \sum_{i=1}^{\nue} W\log_2(1+SINR_{i,\tau} ),
\end{equation}
\vspace{-1em}
and,  using the Jain\textquotesingle s fairness index~\cite{Jain1984}, the users\textquotesingle~fairness as

\begin{equation}
     \label{eqn: fairness}
J_{\tau}^{f} = \frac{(\sum_{i=1}^{\nue} R_{i,\tau})^2}{{\nue} \sum_{i=1}^{\nue} R_{i,\tau}^{{2}}}.
\end{equation}
 The   optimization problem is formulated as the following
\begin{subequations}\label{eq:alloc}
\begin{align}
&\max_{x_{\uav,\tau},y_{\uav,\tau},h_{\uav,\tau},\alpha_{i,\tau}} \omega_r\times R_{\tau}^{\textrm{tot}} +\omega_f\times  J_{\tau}^{f},\\
&\alpha_{i,\tau}> 0,\; \forall \; i \in \{1,..,\nue \},\;  \\
&{\alpha}_{i,\tau}+{\alpha}_{i+1,\tau}=1,\; \forall \; i \in \{1,3,..,\nue-1 \},\;  \\
&R_{i,\tau} \geq R_{min},\; \forall \; i \in \{1,..,\nue \},\; \label{eqn:cstr}\\ 
&L/2 \geq x_{\uav,\tau} \geq - L/2,\; \\
&L/2 \geq y_{\uav,\tau} \geq - L/2,\;\\
& h_{\uav,\tau} \geq h_0,\;  \\
&\omega_r\geq0,\; \omega_f\geq0,
\end{align}
\end{subequations}
where $R_{min}$ is a minimum required rate for each user.
\begin{remark}
To guarantee problem feasibility, 
the power allocation should satisfy:
$        {2^{({-R_{min}}/{W})}} > \alpha_{i,\tau}$, for $i \in \{1,3,..,\nue-1 \}$,  which follows from~\eqref{eqn:cstr}.
\end{remark}
The objective function in~\eqref{eq:alloc} is not convex, hence finding the optimal power allocation and \uav~placement is challenging. Note that an optimal solution should strike a balance between two conflicting objectives: maximizing total users\textquotesingle~data rate and users\textquotesingle~fairness. These two objectives vary drastically based on users-to-\uav~3D distances, environment,  and  spectrum.
Inspired by the recent advances and success of deep reinforcement learning, we propose an efficient framework that
allows the \uav~to  learn   how to maximize objectives,  satisfy requirements, learn  environment patterns,  
and  adapt to related-unseen environments,   
and hence solve efficiently~\eqref{eq:alloc}.

\section{ A\textnormal{dapt}S\textnormal{ky}: Resource Allocation and \uav~Placement Framework}
\label{sec:adaptsky}
\subsection{Framework Preliminaries}

Our reinforcement learning based framework is employed based on the Q learning method. We consider   a sequential decision making setup, where   the \uav~interacts with the network environment which evolves as a Markov process over discrete time steps. 
 At each time step, the \uav~gets a representation of the environment state $s_\tau$,  and  takes an action $a_\tau$ drawn from a set of possible actions according to a certain policy $\pi$. The \uav~moves to a new state $s_{\tau+1}$, and  a reward $r_\tau$ is given as a consequence.  
The aim of the \uav, starting from a time step $\tau$, is to determine the optimal policy,  
which is a series of actions, that maximizes the total discounted reward given by $ \sum_{t=\tau}^{T-1} \gamma^{t-\tau} r_\tau,$
where $\gamma \in [0,1]$ is a discount factor, and $T$ is the total number of time steps.
The point of using a discounted reward   is to make the \uav~give more value to the nearest upcoming rewards. The Q-value of a  state-action pair   $(s_{\tau},a_{\tau})$ is the expected discounted reward obtained from taking action $a_{\tau}$ in state $s_{\tau}$. In the Q-learning method, a Q-table is used to store the Q-values for each state-action pair $Q\left( s_\tau,a_\tau\right)$. 
The computational resources and time required for the iterative process of updating the table in a large state space makes conventional RL techniques inefficient for solving many  optimization problems. DRL methods, nevertheless, like deep  Q-learning (DQL) are emerging to handle  environments represented by large and even continuous state space~\cite{deepmind2013}. 
In DRL,  DNNs are used to  approximate  the  optimal Q-value for a state-action pair observed at $\tau$,  $Q^{\ast }\left( s_{\tau},a_{\tau}\right)$, which is given by Bellman equation as 
\vspace{-1em}
\begin{eqnarray} 
\label{eqn: DRL_Q} Q^{\ast }\left( s_\tau,a_\tau\right) &=&\mathbb{E}\left[ r_{\tau}+\gamma \max_{a_{\tau+1}}Q^{\ast }\left( s_{\tau+1},a_{\tau+1}\right)\right].
\end{eqnarray}
 $Q^{\ast }\left( s_{\tau+1},a_{\tau+1}\right)$ is the optimal Q-value for the next state-action pair ($s_{\tau+1},a_{\tau+1}$). The policy DNN in DQL updates its  parameters $\theta$ every time step $\tau$ with the objective of finding the optimal policy $\pi^{\ast }=\argmax_a Q^{\ast }\left( s_{\tau},a_{\tau}\right)$. This is implemented by minimizing the loss $L(\theta)$, 
   determined by comparing the outputs of
 the policy network $Q\left( s_{\tau},a_{\tau}\right)$  and target network, $r_\tau+\gamma Q^\prime \left( s_{\tau},a_{\tau}\right)$ and given by 
\vspace{-0.5em}
\begin{equation}
   \label{eqn: loss}
   { L(\theta) = \mathbb{E}[\big(r_\tau + \gamma {\max Q^\prime(s_{\tau+1}, a_{\tau+1})} - {Q(s_\tau,a_\tau)}\big)^2]}.
\end{equation}

The target network, which  improves stability of the DQL, has parameters $\theta^{\prime}$ which are cloned with the policy network parameter periodically.
%
To improve the DRL stability even further,  
  DRL randomly, at each time step, samples a mini-batch from an experience replay buffer that stores $s_\tau, a_\tau, s_{\tau+1}$ and $ r_{\tau} $   to calculate the loss and updates the policy network parameters.

%
 
 To enhance the learning process  speed and generalization, we, furthermore, integrate the DN architecture with the DRL. In DN DRL, both target and policy networks have two output layers the value function,  $V(s_{\tau})$, and  advantage function, $A(s_{\tau}, a_{\tau})$ defined as $A(s_{\tau}, a_{\tau}) = Q(s_{\tau}, a_{\tau}) - V(s_{\tau})$. 
%
The value function, defined as the expected value of the Q value,  measures how good it is to be in a given state. The  relative measure of the importance of an action at a particular state is indicated by the advantage function, however.  
\vspace{-1em}
\subsection{\adapt: DRL-based Framework}
Based on the described advanced DRL method, we propose \adapt~a framework 
that allows the \uav~to efficiently position in a 3D plane while efficiently serving the different users. \adapt~helps solve the optimization problem defined in~\eqref{eq:alloc}. Nevertheless, setting up the right learning environment   plays the crucial role in achieving so. Next, we formally describe how the  states, actions, and rewards are designed.  
\begin{enumerate}
\item{\textbf{States.}}
A state $s_{\tau}$ describes the relative locations of the \uav~to each user,   user's power coefficient, and \uav-$\ue_i$ channel gain. $s_{\tau}$ is  defined as 
$
s_{\tau}=\big[s_{1,\tau}^T, s_{2,\tau}^T, ..., s_{{\nue},\tau}^T, h_{\tau}\big]^T$, 
where
$
    s_{i,\tau}=\big[\Delta x_{\uav-i,\tau},\Delta y_{\uav-i,\tau},\alpha_{i,\tau},g_{i,\tau}(d_{i,\tau})\big]^T$, $\forall \; i \in \{1,..,\nue \}$. $\Delta x_{\uav-i,\tau}$ and $\Delta y_{\uav-i,\tau}$ are the step distance between the \uav~and $\ue_i$ in the x-axis and  y-axis respectively, and $h_\tau$ is the \uav~current height. The initial sate, $s_0$, is set based on the predetermined initial \uav~location $(0,0,h_{\uav,{0})}$ and power allocation coefficient $\alpha_{i,0}$ $\forall \; i \in \{1,..,\nue \}$.  The cardinality of $s_{\tau}$   is $4\times \nue+1$. 

\item{\textbf{Actions.}}
An action $a_{\tau}$ is defined as
$
a_{\tau}=[d_x \delta_x,d_y \delta_y,d_h \delta_h, 
\delta_\alpha^T]^T$
 where $\delta_\alpha$ is defined as
$
 [d_{1} \delta_{1},d_{3} \delta_{3}, ...,d_{\nue-1} \delta_{\nue-1}]^T$, $d_{i}\in \{1,-1\}$ and $\delta_{i}$ is the magnitude of change in the power allocation coefficient of   $\ue_i$ $\forall i \in \{1,3,..,\nue-1\}$. 
 $a_{\tau}$ determines the adjustments of the \uav~3D placement and the power allocated to all users. The action vector has a cardinality, denoted by $|A_c|$, of $3+\nue/2$. 
 


\item{\textbf{Rewards.}}
We define the reward at time step $\tau$ as
%
\begin{equation} \label{eq1}
\begin{split}
r_\tau = w_r\times \frac{R_{\tau}^{\textrm{tot}}}{W}   \times \prod_{i=1}^{\nue} \mathds{1}_{ \big(\displaystyle R_{i,\tau}\geq R_{min}\big)} \\
+ w_f\times J_{\tau}^{f} \times \mathds{1}_{ \big(\displaystyle  R_{min}=0\big)} + w_g\times g^{tot}_{\tau}\\
+ w_s \times \sum_{i=1}^{\nue} \mathds{1}_{\big(\displaystyle R_{i,\tau} \geq R_{min}\big)}\\
+ w_u \times \sum_{i=1}^{\nue}\frac{R_{i,\tau}}{W} \mathds{1}_{ \big(\displaystyle R_{i,\tau} < R_{min}\big)} ,
\end{split}
\end{equation}

where 
$g^{tot}_{\tau}$ defines the total users channel gain, $\mathds{1}_{(.)}$ is the indicator function.
$w_r$, $w_f$,  $w_g$,  $w_s$, and $w_u$, with values  greater than or equal to zero,  are the  weights corresponding to  total rate, fairness, total channel gain, and satisfied and unsatisfied minimum rate requirements rewards respectively. The reward is designed carefully in terms of structure weights to make the \uav~learn patterns to solve~\eqref{eq:alloc}. 
The total rate reward term aims to increase the total sum rate after all users meet the minimum rate constraint to ensure fairness. 
The fairness reward term is only relevant when there is no fairness imposed through a minimum rate. It is intended to preclude the \uav~from preferring some users over the rest. 
Without including  such a reward,  the \uav~may end up  favoring some users based on their channel conditions over others given the improvement they offer to the total rate specially in the presence of NLoS links. 
Even though, the impact of the channel gain is implicitly considered in the total rate reward, it makes the \uav~learns channel conditions' patterns and any related spatial variations.
As a way to reinforce the \uav~to satisfy the minimum rate requirement, at any time step $\tau$, a reward of $w_s$, which takes a relatively value,  gets added to the total reward for every  user achieves a rate that exceeds  $R_{min}$. 
Users with a rate lower than $R_{min}$, however, get the unsatisfied minimum rate requirements reward, which aims mainly to encourage the  \uav~to keep improving users rate until the minimum rate requirements is satisfied.

\end{enumerate}

Having described the different states, actions, and rewards, we describe   \adapt~and present it as in Algorithm \ref{alg:adaptsky}. 
At the initialization, all  DN DRL parameters are set.  
    The policy  network    weights and biases ${\theta}$ are initialized randomly, and the target network parameters  $\theta^{\prime}$ are cloned with ${\theta}$.
Similarly, network environment is set and  \uav~location   and power allocation coefficients are initialized. 
%
%
The decision process of \adapt~is made over $E$ episodes with $T$ time steps each. To improve the learning experience at the beginning of each episode the \uav~and power allocation are set back to their initial value. 
%
%
In our time sequential decision process, for a given state \adapt~takes an action from the action space based on the $\epsilon$-greedy algorithm.   To allow for exploring the action space,  at any given time step, the \uav~takes  a random action with a probability   $\epsilon$. Staring from  $\epsilon_s$,  $\epsilon$ gets decayed with a certain rate and converges to an ending value $\epsilon_{e}$. As a way to exploit  the policy network decisions, however,  with a complementary probability, the \uav~takes  the action that maximizes the Q-value which is constructed from the policy network outputs $V$ and $A$ values as $Q(s_{\tau}, a_{\tau} ) \leftarrow  V(s_{\tau}) + A(s_{\tau}, a_{\tau}) - \frac{1}{|A_c|} \sum_{ {a}_{\tau}} A(s_{\tau}, a_{\tau})$. 
After executing $a_\tau$, the \uav~observes the reward $r_\tau$ and the location and power coefficient  to ensure that they are both feasible and modify the resultant state $s_{\tau+1}$ if needed.
After that,  states along with action and the resultant reward are stored in $\zeta$ that has a capacity $M$. Then, a mini-batch of $B$ is sampled from $\zeta$ and used to train the policy network such that the loss $L(\theta)$, found based on   $Q$ and $Q^{\prime}$,  is minimized using the gradient descent algorithm with a learning rate, $l_r$.  $Q^{\prime}$ is constructed from target network output as   $Q^{\prime}(s_{\tau+1}, a_{\tau+1})\leftarrow  V^{\prime}(s_{\tau+1}) + A^{\prime}(s_{\tau+1}, a_{\tau+1}) - \frac{1}{|A_c|} \sum_{{a}_{\tau+1}} A^{\prime}(s_{\tau+1}, a_{\tau+1})$.
Accordingly, the policy notwork parameters $\theta$ are updated      and similarly the target network parameters $\theta^\prime \leftarrow \theta$  after every $\delta$ time steps.  
%
%
 %
\begin{algorithm}[] 
\caption{\adapt}
\label{alg:adaptsky}
\textbf{Initialization:}
Set up DN DRL parameters, and initialize policy  network $Q$ with random  parameters ${\theta}$ and  target network parameters  $\theta^{\prime} \leftarrow \theta$. 
Initialize network environment parameters including users and \uav~locations, and   allocated power \\
 \For{episode ep= $0$:$E-1$}{
  Receive   state $s_0$ \\
\For{$\tau$ = 0:$T-1$}{
Set $\epsilon  \leftarrow \epsilon_{e} + (\epsilon_s - \epsilon_{e}) \times e^{-\frac{(ep \times T+\tau)}{\chi}}$ \\
Take $a_\tau$ based on $\epsilon$-greedy \\
Observe reward $r_\tau$ and next state $s_{\tau+1}$\\
 If a \uav~coordinate or a power  coefficient is infeasible based on $a_\tau$, modify it with a unit magnitude of change and update $s_{\tau+1}$    \\
Store experience ($s_\tau,a_\tau,r_\tau,s_{\tau+1}$) in $\zeta$\\
Set $s_\tau \leftarrow s_{\tau+1}$\\ 
 \If {experience memory size exceeds $B$}
 {Sample random mini-batch  from $\zeta$\\
 Pass the mini-batch to the policy network \\
 Construct $Q(s_{\tau}, a_{\tau})$ from  $V$ \&  $A$ \\

 Pass $s_{\tau+1}$ {mini-batch} to the target network \\
  Construct  target Q-value  $Q^{\prime}(s_{\tau+1}, a_{\tau+1})$    \\ 
 Calculate the loss $L(\theta)$ 
 \\
 Update   $\theta$  such that  $L(\theta)$ is minimized}}
 \If {ep is a multiple of $\delta$} 
 {Update weights and biases of the target network to the weights and biases of the policy network $\theta^\prime \leftarrow \theta$}}

\end{algorithm}

\section{A\textnormal{dapt}S\textnormal{ky} Performance Evaluation}\label{sec: numr}
\subsection{AdaptSky Implementation Details}

 \adapt~networks consist of $2$ fully-connected layers of $128$ neurons each, and use  ReLU as an activation function.  {$B$}  and  $M$ are taken to be $128$ and  $15,000$, respectively.
We employ $\epsilon_s=0.9$ and $\epsilon_{e}=0.1$. $\chi$, used in  $\epsilon$ decaying rate, is set to $200$. Moreover, $l_r$, and $\gamma$ are chosen to be $0.001$ and  $0.999$ respectively.
 We use Adam optimizer~\cite{zhang2018improved}, which is adapted to noisy problems with sparse gradients, to update DNN parameters {according to the value of the $l_r$}.
Target network parameters are updated     every  $10$ episodes.
\adapt~is trained for $E=1000$ episodes with $T=300$ time steps each. In addition to training, we  deploy our \adapt~trained model and test its decision performance over $1000$ time  steps for network environment different than that set for the training.
 \vspace{-1em}
\subsection{Network  Settings}
\label{subsec:Simulation_setup}

 \adapt~performance is tested for a $100 \times 100$ $m^2$ area $\mathcal{A}$ with $\nue =4$ divided into $2$ clusters. Users $\ue_1$ and $\ue_2$  belongs to the same cluster based on their locations. 
The Cartesian coordinates of the users are set to  $(x_1,y_1) = (4,15)$, $ (x_2,y_2) = (-44,-49)$, $ (x_3,y_3) = (-5,21)$ and $(x_4,y_4) = (47,49)$ in \adapt~training scenarios  and uniformly randomly distributed over $\mathcal{A}$ during  the testing.
The minimum \uav~height {$h_{0}$} is set to $ 10$ m.  At the beginning of each {episode}, the initial height of the \uav, {$h_{\uav,0}$}, is set to  $50$ m   and the  power allocation coefficient  $\alpha_{i,0}$ to $0.5$ $\forall i \in \{1,2,..,\nue \}$.
$\delta_x$, $\delta_y$, and $\delta_{h}$ are all  are set to be $1$ m. The magnitude  of  change  in  the  power  allocation is set for $0.01$ for all users. 
All  channel model parameters for both mmWave and \sub~are shown in Table \ref{tab: network parameters}. 
The channel is modeled according to~\cite{akdeniz2014millimeter} which provides the New York city model. 
\begin{table}[h]
    \centering
        \caption{Network parameters value.}
    \label{tab: network parameters}
    \begin{tabular}{l c c}
        \toprule
         \textbf{Parameter} &  \textbf{mmWave} & \textbf{\sub}  \\
         \toprule
         Carrier frequency $f_c$  &28GHz&2GHz\\
         Transmit power $P_T$ &  $20$ dBm & $30$ dBm \\
         Antenna configurations $\mathcal{N}_{\uav} \times \mathcal{N}_{\ue}$ & $8 \times 8$ & $1 \times 1$ \\
         System bandwidth $W$ & $2$ GHz & $50$ MHz \\
         Thermal noise power $\sigma^2 $ & $-84$ dBm & $-88$ dBm \\
          LOS probability parameter $C$  & $9.6117$ & $0.6$ \\
          LOS probability parameter $Y$  & $0.1581$ & $0.11$ \\
          {Path loss} intercept $C_{LoS}$ & $10^{-6.4}$ & - \\
         {Path loss} intercept $C_{NLoS}$ & $10^{-7.2}$ & - \\
         Mean additional LoS  {path loss} $\eta_{LoS}$ & - & $1$ dB \\
         Mean additional NLoS  {path loss} $\eta_{NLoS}$ & - & $20$ dB \\
          {Path loss} exponent $a_{LoS}$ & $2$ & - \\
          {Path loss} exponent $a_{NLoS}$ & $2.92$ & -  \\
         Minimum elevation angle $\theta_0$ & - & $15^{\circ}$ \\
         \toprule
    \end{tabular}
\end{table}

\vspace{-1em}
\subsection{Performance Analysis}
\label{subsec:Numerc_analysis}
In this subsection, we provide the performance  of \adapt~in managing the 3D NOMA-\uav~network  both during the training and testing cases. 
We compare our finding with the state-of-art technique in~\cite{Chen2019}
which throughout the section will be referred to as SoA. The authors in~\cite{Chen2019}, solve the NOMA power allocation and \uav~placement problem using the conventional optimization framework. The \uav~placement in~\cite{Chen2019} is restricted over a 2D plane  which we set its height similar to \adapt~initial height. They furthermore assume users to be only LoS.
For simplicity, we only consider large scale fading and intend to study small scale fading and time-variability in our future work.
For  evaluation, we introduce the performance metrics  ${R}_{e}^{\textrm{tot}}$ and $J_{e}^{f}$ which are defined as achieved average sum-rate and fairness index respectively. ${R}_{e}^{\textrm{tot}}$ ($J_{e}^{f}$) is    determined by averaging the average rate (fairness index) per episode over the most recent $100$ episodes. 
${R}_{e}^{\textrm{tot}}$ and $J_{e}^{f}$ both equal zero for all episodes $ep < 100$.
\subsubsection{Performance of \adapt~in the \sub~spectrum}
\begin{figure}[ht]
    \centering
    \includegraphics[width=0.3\textwidth]{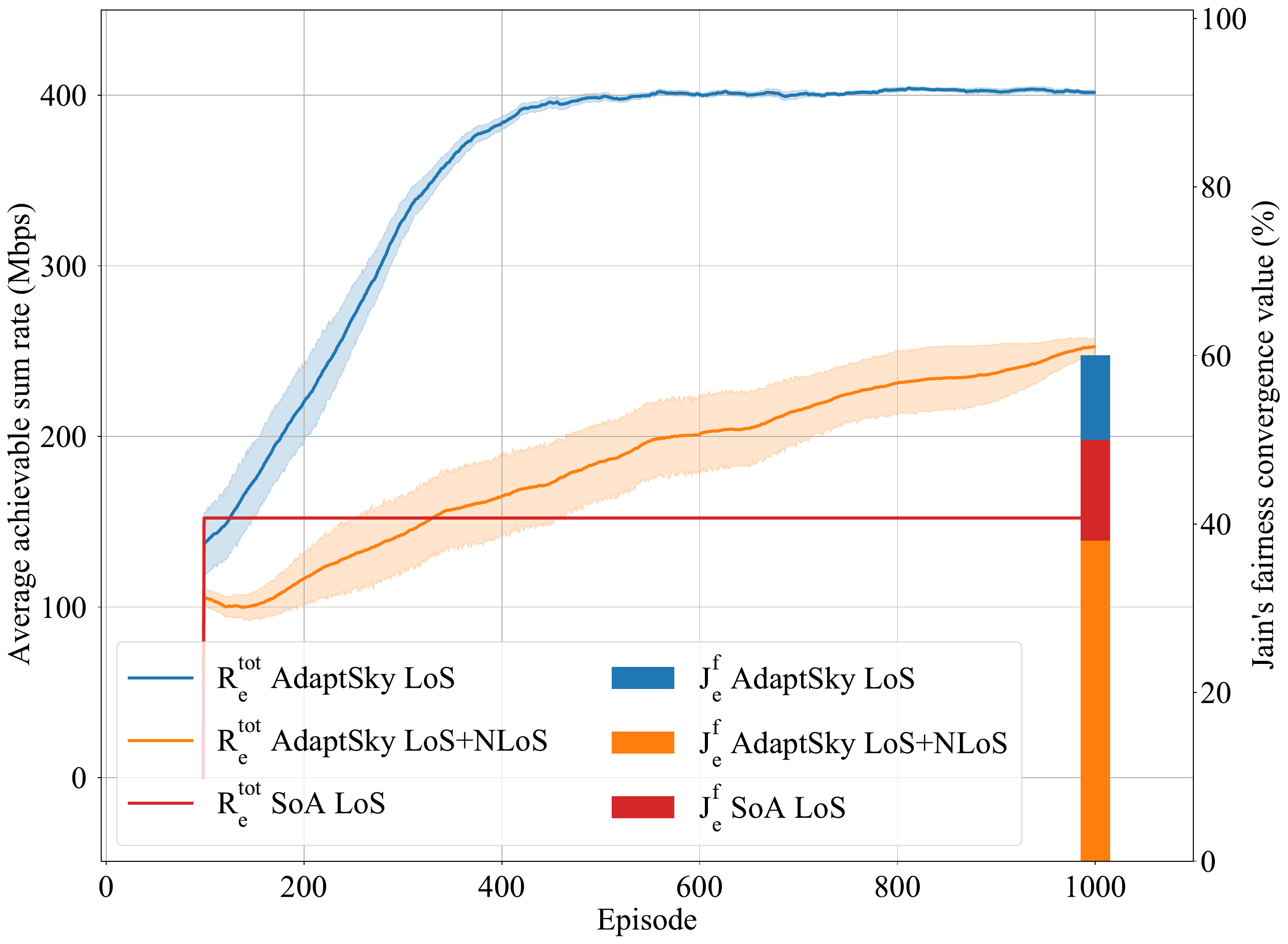}
    \caption{ Average achievable rate versus episode and Jain\textquotesingle s fairness index convergence value in \sub} 
    \label{fig:AdaptSky-Sub-6-ALL-MODELS-s} 
\end{figure}

%
In Fig.~\ref{fig:AdaptSky-Sub-6-ALL-MODELS-s}, we depict ${R}_{e}^{\textrm{tot}}$ achieved by \adapt for both cases where users-\uav~links are dominated by LoS and the case where the channel is generic. We also show the convergence value of the ${J}_{e}^{f}$ at $ep=E-1$.  
In {LoS} scenario, we set $w_r = 1, w_f = 0$, $w_g = 10^7$, $w_s =0$ and $w_u = 0$, with $R_{min} = 0$, while in the generic scenario we set $w_f =5$ so that we prevent the \uav~from favoring one user while ignoring the others (since it is the best way to achieve high sum rate with the existence of NLoS users). The rest of the weights are employed identical to the LoS case.  
The simulations have been conducted for 10 runs and the average and the confidence interval of 1 standard deviation of ${R}_{e}^{\textrm{tot}}$ has been calculated and plotted as shown in Fig.~\ref{fig:AdaptSky-Sub-6-ALL-MODELS-s}. 
\adapt~tends to adapt the power allocation and \uav~placement continuously such that ${R}_{e}^{\textrm{tot}}$ keeps improving until it convergences to a certain value. In our scenario, \adapt~converges to $400$ Mbps, which is $163\%$ higher than that achieved by  SoA.    \adapt~outperforms SoA despite the fact that \adapt~is allocating resources in a more complex environment as the SoA can only place the \uav~in a 2D plane and assumes LoS users.
Moreover, Not only \adapt~achieves a higher   rate performance, but also maintains a $60\%$ fairness index which is more than $20\%$ higher than SoA.
Although  it is not reasonable to compare \adapt~to SoA with the presence of NLoS users since SoA has only LoS users, \adapt~achieves a better ${R}_{e}^{tot}$ and maintains a $40\%$ fairness level  even though the NLoS results approximately in two orders of magnitude worse channel gain. 
\vspace{-1.5em}

\subsubsection{ \adapt~for  mmWave-NOMA-\uav~Networks}  
\begin{figure}[h]
    \centering
    \includegraphics[width=0.3\textwidth]{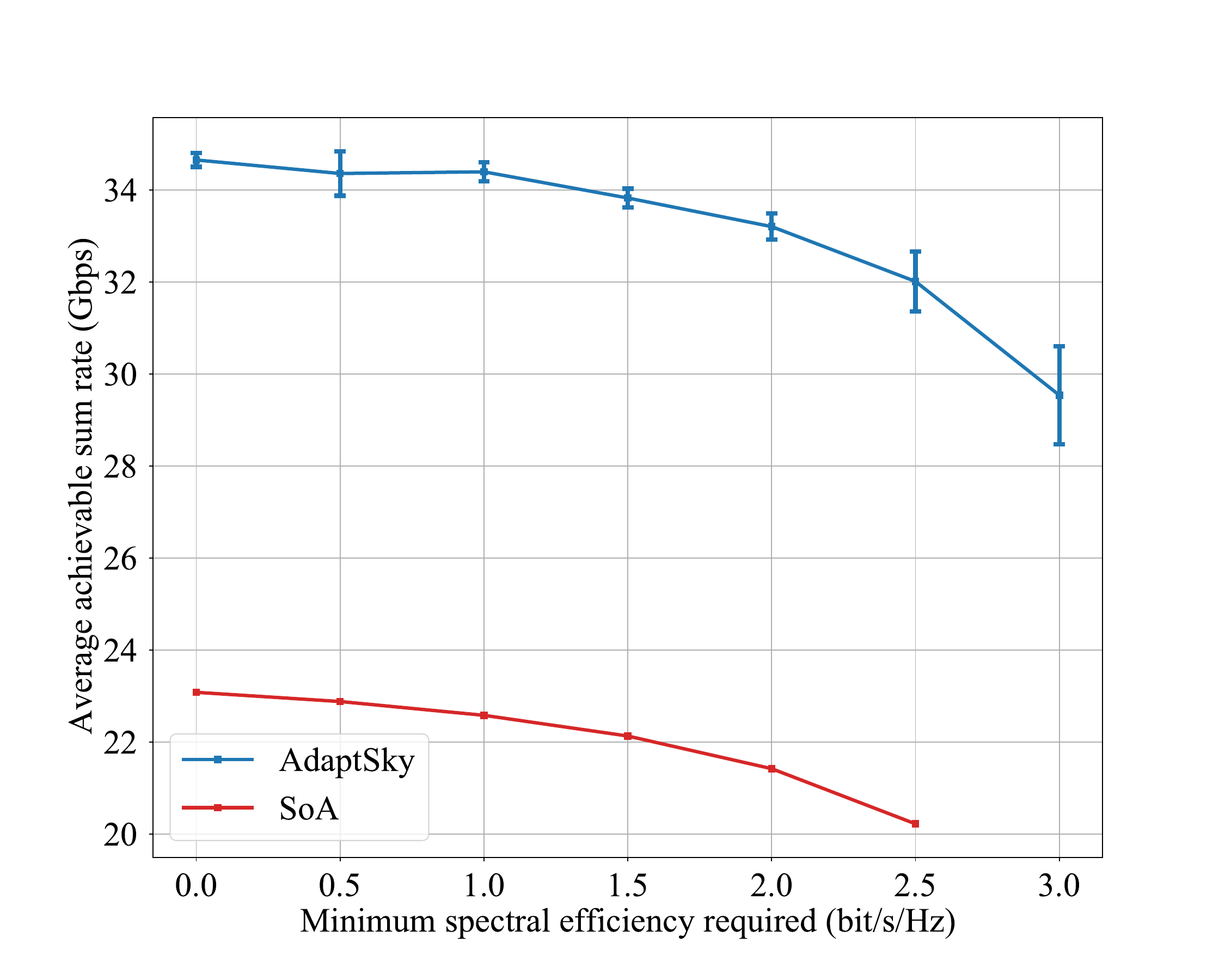}
    \caption{Average achievable sum rate as a function of minimum spectral efficiency required in mmWave.}
    \label{fig: min-sum-vs-se}
\end{figure}

We below evaluate \adapt~performance in terms of training and testing processes for different scenarios.\\
\textbf{Training process.} Assuming that   channels are only LoS, throughout this analysis, we train \adapt~to allocate resources by maximizing total average data rate  while satisfying the  minimum spectral efficiency specified by ${R}_{min}/W$.  To satisfy the stated objective and constraint we set $w_r = 10$, $w_f$ \& $w_g$ to be zero, $w_s$ and $w_u$ to be $100$ and $10$ respectively.  
In Fig. \ref{fig: min-sum-vs-se}, we plot ${R}_{e}^{\textrm{tot}}$ as a function of the minimum spectral efficiency along with the confidence interval of 6 runs.  
Observe that \adapt~has a superior performance than SoA, thanks to its ability to relocate the \uav~in the 3D plane while allocating power at the same time. In addition, \adapt~managed to serve users while providing them with a minimum of  $3$ bit/s/Hz  while SoA could only handle up to $2.5$ bit/s/Hz given the same network resources and hence has a higher  resources management efficiency. 
\vspace{-1.5em}

\begin{figure}[h]
    \centering
    \includegraphics[width=0.3\textwidth]{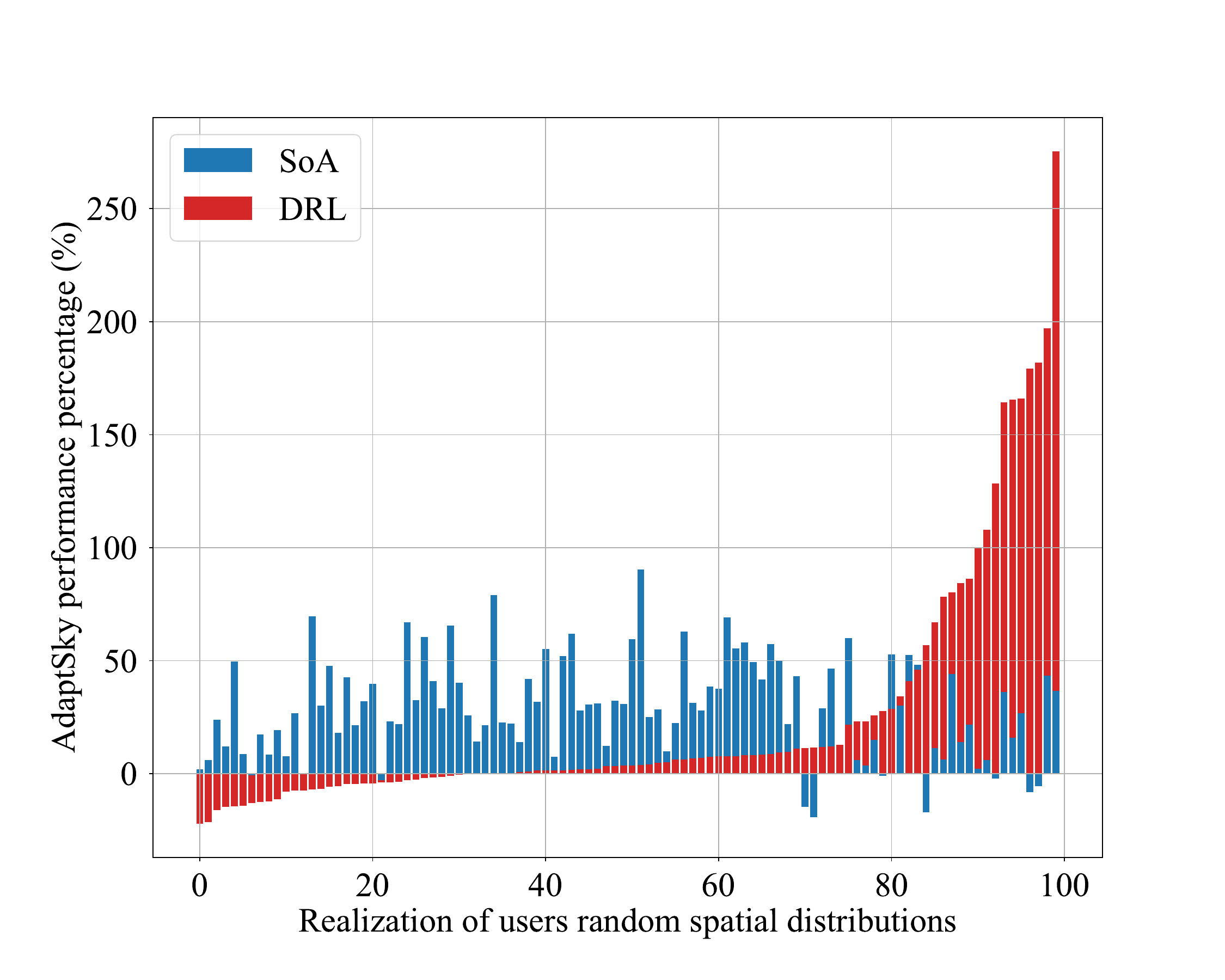}
    \caption{AdaptSky achievable sum rate performance as a percentage of other techniques  for different users location.}
    \label{fig:test-AdaptSky-DN-VS-DRL-VS-SoA}
\end{figure}

\textbf{Testing process.} 
{In this scenario, we   examine  effectiveness, robustness and generalization ability of
\adapt~with respect to both SoA and conventional DRL method. 
We train \adapt~ while setting $R_{min}$ to $0$, $w_r$ to be $10$, and all other  weights to  zero.
We generated $100$ different users location realizations  drawn from a uniform distribution and determined ${R}_{e}^{\textrm{tot}}$ based on  actions taken, based on the trained policy network model, over a $1000$ consecutive time steps.} In 
Fig.~\ref{fig:test-AdaptSky-DN-VS-DRL-VS-SoA}, we draw the average total sum rate percentage  of \adapt~compared with  SoA and DRL. 
\adapt~outperformed the DRL $67\%$ of locations  with average of $39.11\%$ and up to $275.4\%$ improvement.  
This  significant performance improvement  of \adapt~over DRL comes as a result of its higher generalization ability.  Similarly,  \adapt~outperformed SoA over $91\%$ of locations with up to $90.25\%$ improvement and median of $30.25\%$.

%

\vspace{-1.5em}

These results confirm that disjointing the power allocation and the \uav~placement and restricting the \uav~to 2D location  lead to inefficient use of resources. 
Hence, these simulation results demonstrate the significant advantage of our proposed framework over the state-of-the-art, by unleashing the power of AI for solving the power allocation and the 3D-\uav~placement jointly.

\section{Conclusion}
\label{sec: conc}
In this paper, we  proposed \adapt, a novel AI-based framework built based on DRL with DN architectures. 
\adapt~optimizes a \uav~3D location while allocating resources effectively in NOMA-\uav~networks, simultaneously, for a  generic-realistic channel gain for both \sub~as well as mmWave technologies.  
Simulation results showed that  \adapt~yields significant performance gains over conventional approaches in terms of average achievable sum rate and fairness. Moreover, \adapt~shows significant improvement in generalization over the  DRL method.

\bibliographystyle{unsrt}
\bibliography{AdaptSky}

\end{document}